\documentclass[12pt,preprint]{aastex}
\begin{document}

\title{\large A Late-Time Flattening of Afterglow Light Curves}

\author{D. A. Frail\altaffilmark{1}, B. D. Metzger\altaffilmark{2}, E.
  Berger\altaffilmark{3}, S. R. Kulkarni\altaffilmark{3}, S. A.
  Yost\altaffilmark{3}}

\altaffiltext{1}{National Radio Astronomy Observatory, P.O. BOX `O', Socorro,
NM 87801}

\altaffiltext{2}{University of Iowa, Department of Physics and
Astronomy, Iowa City, IA 52242}

\altaffiltext{3}{Division of Physics, Mathematics and Astronomy,
  105-24, California Institute of Technology, Pasadena, CA 91125}

\begin{abstract} 
  We present a sample of radio afterglow light curves with measured
  decay slopes which show evidence for a flattening at late times
  compared to optical and X-ray decay indices. The simplest origin
  for this behavior is that the change in slope is due to a jet-like
  outflow making a transition to sub-relativistic expansion. This can
  explain the late-time radio light curves for many but not all of the
  bursts in the sample. We investigate several possible modifications
  to the standard fireball model which can flatten late-time light
  curves. Changes to the shock microphysics which govern particle
  acceleration, or energy injection to the shock (either radially or
  azimuthally) can reproduce the observed behavior. Distinguishing
  between these different possibilities will require simultaneous
  optical/radio monitoring of afterglows at late times.
\end{abstract}


\section{Introduction}\label{sec:intro}

One of the defining characteristics of X-ray and optical afterglows is
the observed power-law decay of their light curves. On timescales of
hours to days after a burst the exponent $\alpha$ (defined by
$F_\nu\propto t^{\alpha}$) typically lies in the range of $-1$ to
$-2$.  The light curves of several GRBs have also been seen to undergo
an achromatic break, steepening their temporal decay by
$\Delta\alpha\sim{1}$. The standard afterglow model provides a
framework in which to interpret these power-law slopes and their
changes. Synchrotron emission is produced in a relativistic shock
which accelerates electrons to a power-law distribution with energy
index $p$ given by N($\gamma_e)\propto\gamma_e^{-p}$ above some
minimum energy $\gamma_m$. The evolution of the expanding blast wave
is sensitive to the geometry of the shock, the kinetic energy in the
shock, and the structure of the circumburst medium. This leads to the
temporal (and spectral) evolution of the light curves whose power-law
indices are predicted for different cases (Sari, Piran \& Narayan
1998\nocite{spn98}; Sari, Piran \& Halpern 1999\nocite{sph99};
Chevalier \& Li 2000\nocite{cl00}).  The most commonly accepted
explanation for the sharp breaks in optical and X-ray light curves is
that the outflows are collimated and that the change in $\alpha$ is
the result of the Lorentz factor $\Gamma$ dropping below
$\theta_j^-1$, the inverse opening angle of the jet \citep{rho99}.
However, it should be noted that the jet signature is not identified
unambiguously in all bursts, and for some events there are alternate
explanations for the origin of the break \citep{kp00b,mpp+01,wl02}.

The situation at radio wavelengths is slightly more complex.
Initially, at least, the emission comes from the lowest energy
electrons at $\gamma_m$ which are radiating only a small fraction of
their energy at radio frequencies below $\nu_m$.  This gives rise to
the familiar spectral slopes of $\nu^{1/3}$ and $\nu^2$ in the
optically thin and thick part of the radio spectrum, respectively, and
it produces light curves that rise with time. A jet break is expected
initially to produce only a shallow power-law decay (e.g. t$^{-1/3}$
to t$^0$) of the radio light curve until a time when the synchrotron
peak $\nu_m$ passes through the band. If the expansion remains
relativistic, the post-jet power-law decay $\alpha_R$ will be the same
as that at optical and X-ray wavelengths ({\em i.e.,} $\alpha\sim -2$)

In addition to the late onset of the steep jet decay, it may be
possible to use the long-lived nature of the radio emission to look
for changes in $\alpha$ when the expanding shock slows to
subrelativistic speeds (Waxman, Kulkarni \& Frail 1998\nocite{wkf98}).
Deviations in the power-law decay could also be produced by density
enhancements in the circumburst medium on parsec scales, similar to
that seen in supernovae \citep{mwv+00}.  Likewise, it is possible that
the bulk of the GRB blast energy is in a low $\Gamma$ component and
that the late-time afterglow will be refreshed as this slower shock
catches up to the deccelerating swept-up shell.  There may also exist
an additional low energy electron component
\citep{wax97a,bhat01,lc01,pan01}, which while energetically
unimportant it could significantly modify the late-time afterglow
light curves. Finally, all of these processes could be masked by the
emergence of an underlying host galaxy, whose radio emission is the
result of prodigious star formation.

Motivated by the possibility of detecting these effects at late times,
we have analyzed a set of well-sampled GRB radio light curves for
which it is possible to measure their temporal decay indices. In
\S\ref{sec:xo} we compare the sample of X-ray and optical afterglows,
while in \S\ref{sec:ro} the comparison is carried out with the radio
and optical afterglow sample. In \S\ref{sec:select} we discuss known
sources of biases that affect decay slope measurements, while in
\S\ref{sec:nr} and \S\ref{sec:dis} we discuss possible explanations
for the observed flattening of the radio light curves.

\section{An X-ray/Optical Comparison}\label{sec:xo}

In Table \ref{tab:decay} we provide a list (complete until 2002
January) of all afterglows with accurately measured temporal decay
indices at both X-ray and optical wavelengths. Most of the X-ray
measurements were obtained between 8 and 24 hours after the burst. For
the optical data we list the time interval over which the power-law
fit was made. If a jet break occurred in the light curve, two values
of $\alpha_o$ are given along with the relevant timescale. When
required, a constant emission component from a host galaxy was also
fit to the optical light curves.

In the top panel of Figure \ref{fig:alphas} we show a plot of
$\alpha_x$ versus $\alpha_o$. Most of the points are distributed about
the line $\alpha_x=\alpha_o$ with small scatter but there are a few
significant deviations where the X-ray decay is steeper than the
optical decay. For reasonable physical parameters for the shock in a
constant density medium, the synchrotron cooling frequency $\nu_c$ is
expected to lie between the optical and X-ray bands on these
timescales \citep{pk01}. The temporal slope of light curves measured
above $\nu_c$ will be steeper by an amount $\Delta\alpha$=1/4 (Sari et
al.~1998)\nocite{spn98}, consistent with all points which lie above
the fudicial $\alpha_x=\alpha_o$ line.

More than half of the points in Fig.~\ref{fig:alphas} have a tendency
to lie below the $\alpha_x=\alpha_o$ line. In the simplest afterglow
models $\alpha_o$ is not expected to be steeper than $\alpha_x$. We
could expect such a result if the shock wave were propagating into a
density gradient, such as that produced by mass loss from a progenitor
star. For reasonable physical parameters $\nu_c$ is expected to lie
below the optical band for the timescales of interest here
\citep{cl01}. A more plausible explanation is that inverse Compton
emission, which has been reported for several bursts (e.g., Harrison
et al. 2001\nocite{hys+01}), is producing a slight flattening the
X-ray light curves.  Finally, for GRB\,980519 it is likely that the
jet break occurred close to the time of the X-ray measurements
\citep{naa+99,jhb+00} producing an estimate of $\alpha_x$ that is
intermediate between the two asymptotic values.

\section{A Radio/Optical Comparison}\label{sec:ro}

\begin{figure}
\epsscale{0.8}
\plotone{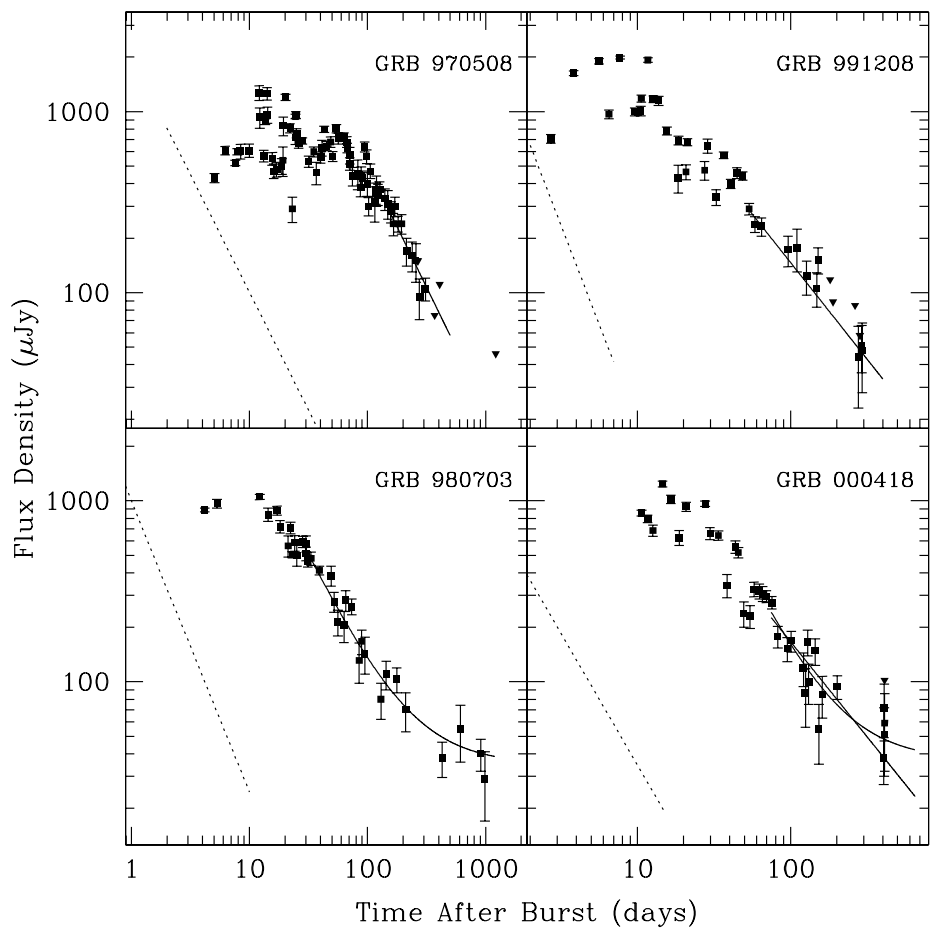}
\caption{Radio light curves of four GRBs at a frequency of 8.46
  GHz. A least squares fit was made to the decaying portion of the
  light curve. The time interval and the slope of each fit is
  indicated by the solid lines. For GRB\,980703 a constant flux
  density was added to the fit to account for the emission from an
  underlying host galaxy. For GRB\,000418 a pure power-law fit is
  shown along with a host component included. The slopes of the
  optical light curves and the time interval over which the fits were
  made are illustrated schematically with dashed
  lines. \label{fig:four}}
\end{figure}

The 10 afterglows in Table \ref{tab:decay} were selected from Frail et
al. (2003)\nocite{fkbw03} to have well-sampled light curves at a
frequency of 8.5 GHz. The remaining 15 GRBs in this catalog do not
have sufficient measurements to determine a decay index $\alpha_R$ at
late times. A least-squares fit was made to each dataset. In order to
be as free as possible from interstellar scintillation and deviations
from power-law decay due to curvature effects in the synchrotron
spectrum (see \S\ref{sec:select}), each burst was fit to several
different starting epochs. The final values of $\alpha_R$ listed in
Table \ref{tab:decay} were often a compromise between making the
starting epoch as late as possible while including sufficient data
points for a fit. In Figure \ref{fig:four} we show the results of our
fits for the four best sampled events. The reduced $\chi^2$ ranges
from 0.6 to 1.3, typical of the sample as a whole.

\begin{figure} 
\epsscale{1.2}
\plottwo{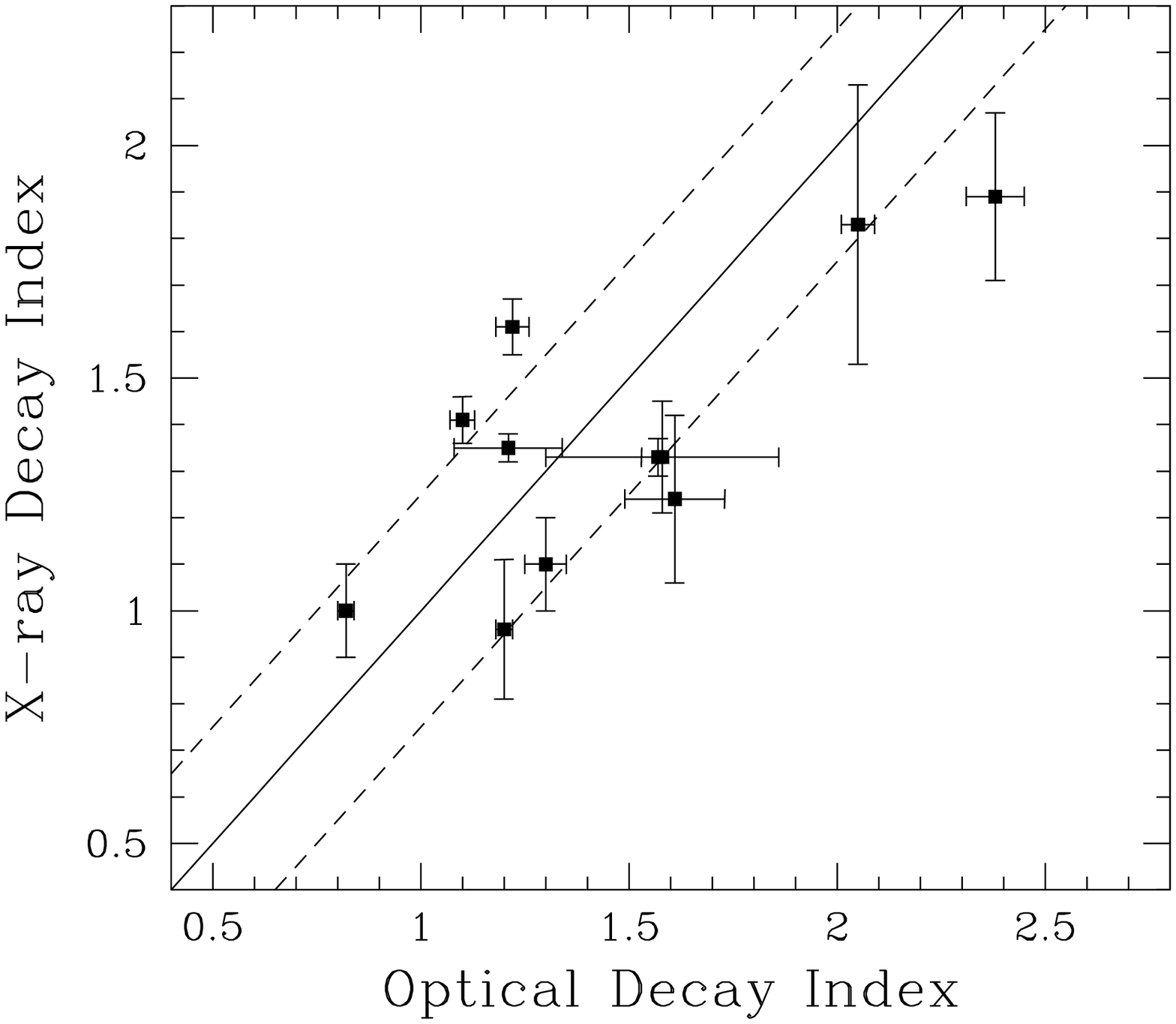}{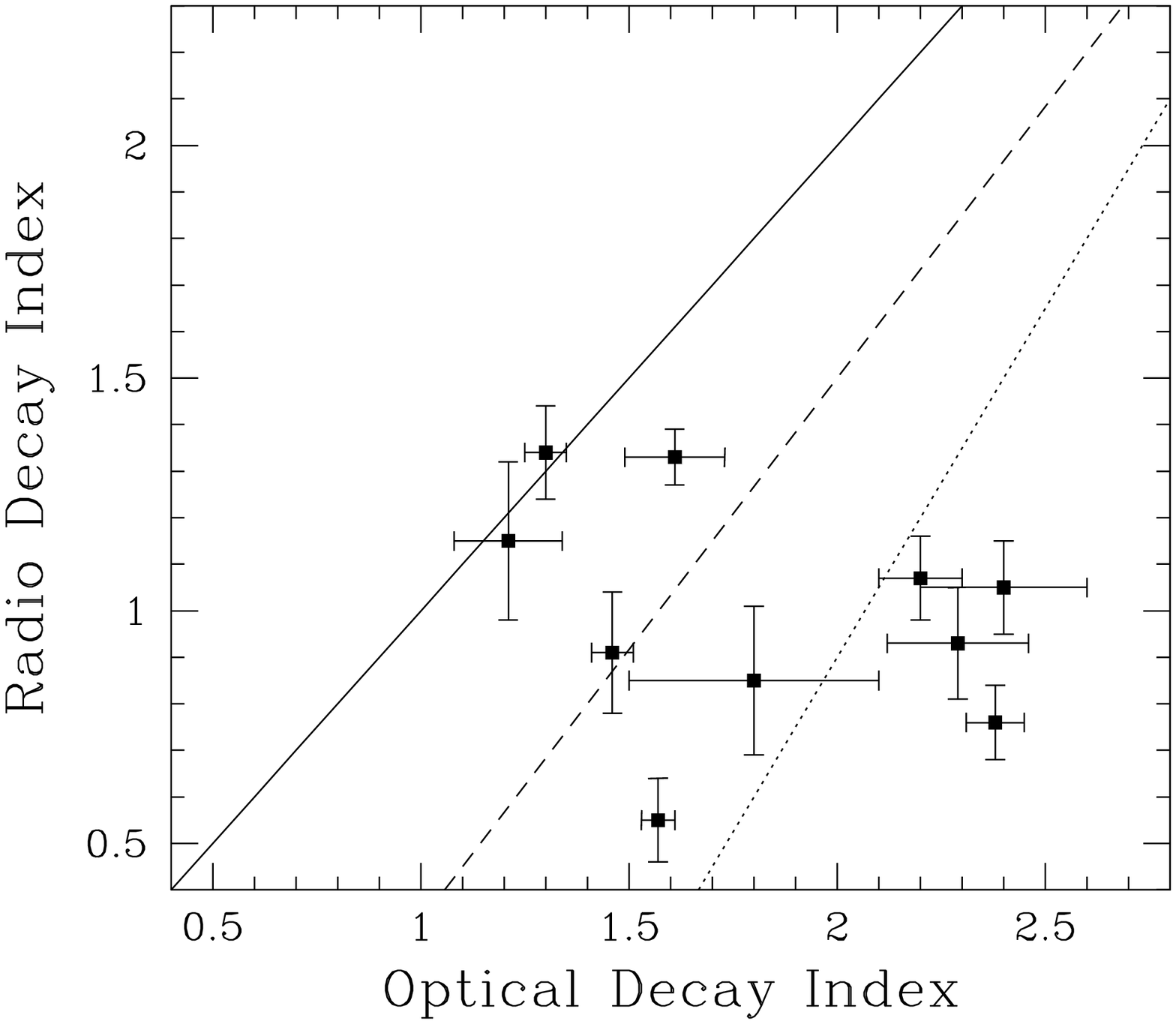}
\caption{The top panel shows the decay indices $\alpha$ defined by
  F$_\nu\propto{t}^{-\alpha}$ for a sample of gamma-ray bursts with
  X-ray and optical afterglows. The solid line corresponds to
  $\alpha_x=\alpha_o$, while the dashed lines have constant offsets of
  $\pm$0.25. The bottom panel shows decay indices for a sample of
  bursts with radio and optical afterglows. The solid line is
  $\alpha_R=\alpha_o$ while the dotted and dashed lines are the
  expected relation between the relativistic and non-relativistic
  temporal decay slopes for a jet-like outflow in constant density and
  wind circumburst media, respectively. In plotting these lines it is
  assumed that $\alpha_o=-p$ and $\alpha_{NR}=\alpha_R$.
\label{fig:alphas}}
\end{figure}

In the bottom panel of Figure \ref{fig:alphas} we show a plot of
$\alpha_R$ versus $\alpha_o$. The optical decay index was chosen over
$\alpha_x$ for this comparison because there are more joint
radio/optical datasets and because their lightcurves are taken over
similar timescales. Nonetheless, for the discussion that follows it is
important to realize that the overlap between the optical and radio
time intervals is relatively small. This will complicate attempts to
understand the relation between $\alpha_o$ and $\alpha_R$ (see
\S\ref{sec:dis}).

There is a clear trend in Figure \ref{fig:alphas}, namely the radio
decay indices are equal to or substantially flatter than the optical
(or X-ray) decay indices.  Moreover, the $\alpha_R$ values rarely
exceed $-1$ and there are no examples of $\alpha_R\sim -2$. The data
can be grouped into three different regions in Figure
\ref{fig:alphas}. The first are those with $\alpha_R\simeq\alpha_o$,
the second have $\alpha_o<-2$, and the third group have $\alpha_o>-2$.
In the next several sections we discuss several possible explanations
for the origin of these temporal slopes.

\section{Systematic Sources of Light Curve Flattening}\label{sec:select}

Some of the measured $\alpha_o$ values are not as steep as is expected
for a jet break due to the presence of a host galaxy, which for most
events dominates the optical light curves typically between one week
and one month after the burst. In those cases, contamination from the
host galaxy may prevent a true determination of the post-jet
$\alpha_o$.  Radio data have been used in several cases (e.g.,
GRB\,000418 and GRB\,980703) to further substantiate claims of jet
breaks \citep{bdf+01,fyb+03} but removing the host galaxy contribution
is often difficult. A second source of flattening for optical light
curves may be the excess of optical flux at $\sim$20(1+$z$) days,
commonly attributed to the rise of a supernova component (e.g. Bloom
et al.~2002\nocite{bkp+02}).

A host galaxy may also flatten the late-time radio light curves.  For
GRB\,980703 there is good evidence that the fit requires an additional
contribution from an underlying galaxy (Berger, Kulkarni \& Frail
2001)\nocite{bkf01}. A starburst host galaxy has also been proposed
for GRB\,000418 \citep{bck+03}. However, while the submillimeter
detection of the host is robust, in our view further multi-frequency
radio observations are required before the centimeter identification
of a host can be considered certain. Adding a host component to the
GRB\,000418 fit yields $\alpha_R=-1.72\pm 0.09$ and $f_{host}$=37
$\mu$Jy.

We do not expect radio galaxies to be detected as frequently as those
at optical wavelengths. In order to significantly flatten the
$\alpha_R$ values in Table \ref{tab:decay} we require F$_{host}\sim
30-40$ $\mu$Jy, which at 8.5 GHz for the typical redshifts of these
events, requires SRF$\sim 10^3$ M$_\odot$~yr$^{-1}$, or
L$_{bol}>10^{12}$ L$_\odot$. If GRBs trace star formation in the
universe, as is currently believed, then only about 20\% of GRBs are
expected to be found in such ultraluminous host galaxies
\citep{bbt+03}. This fraction is in good agreement with a more direct
determination from centimeter and submillimeter observations of GRB
host galaxies \citep{bck+03}. Thus we expect that one, perhaps two,
afterglows in our sample have undetected host galaxies. Fortunately,
continued multi-frequency radio monitoring can be used to detect and
subtract off any galaxy contribution.

A second, more insidious, problem is to ensure that $\alpha_R$ is not
measured when the light curve is in transition between two power-law
behaviors. For example, in the case a jet-like outflow there is a
timescale $t_j$ when the edge of the jet becomes visible and the
optical emission steepens to $\alpha_o=-p$. In contrast at this time,
the radio flux (which is typically emitted below the synchrotron peak
$\nu_m$ at this time) decays with $\alpha_R=-1/3$ until $\nu_m$ passes
through the radio band, after which it decays as $\alpha_R=-p$ in the
relativistic phase or as $\alpha_R=(21-15p)/10$ in the
sub-relativistic phase. Fits which encompass any of these transitions
will give values of $\alpha_R$ which are in between these extremes.

An artificial flattening of the light curve will be measured if the
fit is made before $\nu_m$ passes through the radio band. If $\nu_m$
passes through the optical band at a time $t_o$ it is straightforward
to show that for a jet $\nu_m$ passes through 8.5 GHz at
$t_R\simeq{230}t_j^{1/4}\times{t_o}^{3/4}$ days provided
$t_o<t_j<t_R$.  Thus for typical shock parameters we expect
$t_R\sim5-30$ days (Sari et al.~1999\nocite{sph99}). Most of the fits
in Table \ref{tab:decay} were made at a time $t>t_R$ but we caution
the reader that the shallow $\alpha_R$ values for at least two GRBs
may be the result of fitting over this transition. The only way to
improve on these fits is to reduce the dependence on earlier
measurements through deep observations of the faint afterglow emission
at late times.

%In the previous section we have made the assumption that the measured
%decay indices $\alpha_o$ and $\alpha_R$ are representative of their
%expected asymptotic limits. Thus, for example, in the case of the
%Jet+ISM model we have assume $\alpha_o$=$p$ and
%$\alpha_R$=$\alpha_{NR}$. In this section we will discuss known
%effects that act to flatten optical and radio light curves.

\section{A Dynamical Origin for the Late Decay Slopes}\label{sec:nr}

A change in the temporal slope is expected in the basic afterglow
model when the expansion of the blast wave becomes subrelativistic.
This dynamical transition has been predicted for some time (Wijers,
Rees \& M\'esz\'aros 1997\nocite{wrm97}), and it has been claimed to
have been seen in a number of events (Frail, Waxman \& Kulkarni
2000\nocite{fwk00}; Dai \& Lu 2000\nocite{dl00}; Piro et al.
2001\nocite{pgg+01}, Int'Zand et al.~2001\nocite{zka+01}). This occurs
on a timescale when the rest mass energy swept up by the expanding
shock becomes comparable to the initial kinetic energy of the ejecta.
For kinetic energies of 10$^{51}$ and circumburst densities of 1
cm$^{-1}$ this occurs on a timescale of order 100 days. After this
time the dynamical evolution of the shock is described by the
Sedov-Taylor solutions rather than the relativistic formulation of
Blandford \& McKee (1976\nocite{bm76}).

Independent of geometry, the expected temporal slope in the
non-relativistic regime is $\alpha_{NR}=(21-15p)/10$ for a constant
density (ISM) medium, and $\alpha_{NR}=(5-7p)/6$ for a wind-blown
medium ({\em i.e.,} $\rho\propto{r}^{-2}$). These values assume that
the synchrotron break frequency $\nu_m$ has passed through the band
but the cooling frequency $\nu_c$ remains above the band. Livio \&
Waxman (2000)\nocite{lw00} provide a convenient table of $\alpha$'s
for different cases.

For a spherical fireball undergoing a transition to non-relativistic
expansion, the light curves are expected to {\it steepen} by
$\Delta\alpha\simeq 0.3$, while for a jet-like expansion the light
curves are expected to {\it flatten} by $\Delta\alpha\simeq 1$. By and
large the $\alpha_R$ values in Table \ref{tab:decay} are flatter than
the $\alpha_o$ (and $\alpha_X$) values measured at earlier times. This
immediately rules out a spherical geometry for the majority of bursts
in Table \ref{tab:decay} with measured $\alpha_R$ values. For a
jet-like outflow it is expected that $\alpha_o\simeq{-p}$ (for
$t>t_{jet}$) and therefore the values of $\alpha_{NR}$ can be
calculated and compared to $\alpha_R$. For bursts with
$\alpha_o\simeq{-2.2}$ (GRB\,991208, GRB\,00031C, GRB\,000418 and
GRB\,000926) we expect $\alpha_{NR}\simeq{-1.2}$ in the ISM model and
$\alpha_{NR}\simeq{-1.7}$ in the wind model.  From
Fig.~\ref{fig:alphas} we note that at least for these GRBs the JET+ISM
model provides a better (but not ideal) fit to the observed $\alpha_R$
values than a JET+WIND model. The deviations are in the sense that the
observed $\alpha_R$ are systematically more shallow than expected from
the $\alpha_o$ measurements. This is not likely to be a systematic
bias because it is in the opposite sense of what would be expected if
the optical decay was still in transition to its asymtoptic value
({\em i.e.,} $\alpha_o>{-p}$). The JET+ISM model may also explain
events like GRB\,980703 for which there is good evidence that the
temporal slope $\alpha_o$ is underestimated because the optical light
curves are contaminated by an underlying host galaxy
(\S\ref{sec:select}).

A simple jet model does not work for GRB\,970508, for which
$\alpha_x>\alpha_o\simeq\alpha_R$. We note that both the optical and
radio temporal slopes were measured at comparatively late times and
therefore {\it both} of the fits may be dominated by points when the
afterglow was in the sub-relativistic phase. In this case a
near-spherical shock with $p\simeq{2.3}$ provides a good description
of the light curves (Frail et al. 2000, but see Chevalier \& Li
2000\nocite{cl00}). The most difficult challenge for the JET+ISM model
comes from the third cluster of GRBs in Fig.~\ref{fig:alphas} with
$\alpha_o>-2$. In two cases (GRB\,010222, and GRB\,000911) adopting
$\alpha_o\simeq{p}$ yields estimates of $\alpha_{NR}$ that are
substantially flatter than the observed $\alpha_R$ values. The
JET+WIND model works better but it is not clear that this model can
explain the pre-jet behavior of the light curves.

To summarize, while a sub-relativistic transition of a jet-like
outflow in a constant density medium provides a reasonable explanation
for the flattening of the radio light curves for some of the GRBs
significant departures from this behavior are seen (e.g. GRB\,000926).
The radio slopes are substantially flatter than expected in this
simple model and some other source must be found.

\section{Alternate Origins and Conclusions}\label{sec:dis}

In the previous sections we have presented evidence that the temporal
decay of radio light curves measured months after the burst show
evidence for a flattening compared to the decay slopes measured at
optical and X-ray wavelengths within the first week after the burst.
The simplest explanation, consistent with most of the data, is that
the flattening is the result of a dynamical transition of the shock to
sub-relativistic expansion.

There are other physical effects that may lead to the flattening of
radio lightcurves. Several authors (Bhattacharya 2001\nocite{bhat01};
Li \& Chevalier 2000\nocite{lc00}; Panaitescu 2001\nocite{pan01};
Dado, Dar \& De R\'ujula 2003\nocite{ddr03}) have modified the
electron energy distribution by introducing a break in the power-law
below which the spectrum is hard ({\em i.e.}, $p<2$).  This has the
virtue of requiring no additional source of energy while
simultaneously fitting for the different decays slopes in the X-ray,
optical and radio bands.  Its disadvantages are that it requires the
introduction of another free parameter in the modeling, and that
current simulations of particle acceleration in ultrarelativistic
shocks are unable to produce hard energy spectra \citep{agkg01}. Other
modifications to the shock microphysics are possible \citep{rr03}. For
example, Yost et al.~(2003)\nocite{yost03} have relaxed the usual
assumption that the magnetic energy density behind the shock is
constant, and note that the late-time light curves flatten when the
magnetic energy grows inversely with the Lorentz factor of the shock.

The flattening could also be maintained by a continuous or episodic
injection of energy, rather than the one-time injection as is commonly
assumed \citep{rm98,sm00,kp00c}. Slower moving shells of ejecta catch
up to the deccelerating main shock and re-energize it, causing to
afterglow to brighten at all wavelengths (e.g.~Panaitescu,
M\'esz\'aros \& Rees~1998\nocite{pmr98}). This ``refreshed shock"
model has been invoked to explain the optical behavior of GRB\,010222
\citep{bhpf02} and GRB\,021004 \citep{fyk+03}. This explanation is
problematic for the radio flattening because it requires continuous
energy injection but with a delayed turn-on in order to both maintain
the shallow decay of the radio light curve, while preserving the
steeper optical decay. One way to overcome this difficulty would be to
add energy though a two component jet-like outflow, with the radio
emission originating from an outflow with a wide opening angle
carrying the bulk of the energy. This was first proposed for
GRB\,991216 \citep{fbg+00}, but Berger et al.~(2003)\nocite{bkp+03}
recently have made a stronger case for GRB\,030329. Since the bulk of
the energy in this case is carried by the slower moving ejecta a
crucial test would be to carry out a minimum energy analysis like for
GRB\,970508 (Frail et al.~2000)\nocite{fwk00} and look for large
excesses compared to energies derived from conventional methods
\citep{pk02,fks+01}.

In principle it should be possible to distinguish between these
alternate explanations for the observed flattening. A dynamical
transition is achromatic, and so a break in the optical and radio
light curves should occur at the same time. A modified electron
spectrum will produce two spectral components, recognized by comparing
the spectral slopes both within and between the radio and optical
bands. Energy injection, added either radially via refreshed shocks,
or azimuthally from complex jet structure, will most likely be
identified from multi-component light curves. All of these tests
require near-simultaneous optical and radio light curves but as is
evident from Table \ref{tab:decay} such data is currently lacking.
Although there are several practical problems to overcome
(\S\ref{sec:select}), future monitoring of optical afterglows should
be extended to produce a better overlap with the late-time radio
measurements. The recent bright and nearby GRB\,030329 is a promising
candidate.
% duh!

\acknowledgements

The National Radio Astronomy Observatory is a facility of the National
Science Foundation operated under cooperative agreement by Associated
Universities, Inc. DAF thanks the Astronomical Institute at the
University of Amsterdam for their hospitality during the time when
this paper was being written.

%\bibliographystyle{apj1b}
%% This is what I use here at the AOC
%%\bibliography{journals_apj,slopes,../g021004/radio/mastergrb}
%\bibliography{journals_apj,slopes,mastergrb}

\clearpage

\begin{deluxetable}{ccccccc}
\tabletypesize{\small}
\tablecolumns{7}
\tablewidth{0pc}
\tablecaption{X-ray, Optical and Radio Decay Slopes\label{tab:decay}} 
\tablehead {
\colhead {}             &
\colhead {}             &
\colhead {}             &
\colhead {Epoch}        &
\colhead {}             &
\colhead {Epoch}        &
\colhead {Ref.}         \\
\colhead {GRB}          &
\colhead {$\alpha_X$}   &
\colhead {$\alpha_o$}   &
\colhead {(days)}       &
\colhead {$\alpha_R$}   &
\colhead {(days)}       &
\colhead {}             
}
\startdata
% GRB  &  alphax           & alpha_o       & Epoch & alphaR & epoch & refs
970228  & $-1.33 \pm 0.12$ & $-1.58 \pm 0.28$ & 1-5   & \nodata & \nodata & 1,2 \\ 
970508  & $-1.1 \pm 0.1^a$ & $-1.30 \pm 0.05$ & 2-120 & $-1.34 \pm 0.10$ & 115-309 & 3,4 \\ 
971214  & $-0.96 \pm 0.15$ & $-1.20 \pm 0.02$ & 0.5-3 & \nodata & \nodata & 5,6 \\ 
980329  & $-1.35 \pm 0.03$ & $-1.21 \pm 0.13$ & 0.7-10& $-1.15 \pm 0.17$ & 55-121 & 7,8 \\ 
980519  & $-1.83 \pm 0.3$  & $-2.05 \pm 0.04$ & 0.35-2& \nodata & \nodata & 9,10 \\     
980703  & $-1.24 \pm 0.18$ & $-1.61 \pm 0.12$ & 0.8-10& $-1.33 \pm 0.06^b$ & 27-210 & 5,11 \\   
990123  & $-1.41 \pm 0.05$ & $-1.10 \pm 0.03$ & 0.18-2& \nodata & \nodata & 5,12 \\     
990510  & $-1.0  \pm 0.1$  & $-0.82 \pm 0.02$ & 0.15-1.2& \nodata & \nodata & 13,14 \\     
991208  &     \nodata      & $-2.2  \pm 0.1$  & 2-7   & $-1.07 \pm 0.09$ & 53-293 & 15\\
991216  & $-1.61 \pm 0.06$ & $-1.22 \pm 0.04$ & 0.4-2 & \nodata & \nodata & 16,17 \\      
        &     \nodata      & $-1.80 \pm 0.30$ & 2-15  & $-0.85 \pm 0.16$ & 8-78 &  \\
000301C &     \nodata      & $-2.29 \pm 0.17$ & 4-13  & $-0.93 \pm 0.12$ & 45-165 & 18 \\
000418  &     \nodata      & $-1.41\pm0.08^c$ & 2-14  & $-1.05 \pm 0.10$ & 75-202 & 19 \\
000911  &     \nodata      & $-1.46 \pm 0.05$ & 1-15  & $-0.91 \pm 0.13$ & 3-23 & 20 \\
000926  & $-1.89 \pm 0.18$ & $-2.38 \pm 0.07$ & 2-30  & $-0.76 \pm 0.08$ & 25-288 & 21,22 \\
010222  & $-1.33 \pm 0.04$ & $-1.57 \pm 0.04$ & 1-50  & $-0.55 \pm 0.09$ & 5-206 & 23,24 \\
\enddata
\tablecomments{\small The columns are (left to right): (1) GRB name,
  (2) the X-ray temporal decay index defined by $F_X\propto
  t^{\alpha_X}$, (3) the optical temporal decay index, (4) the
  timerange over which the power-law fit was made to the optical light
  curves, (5) the radio temporal decay index, (6) the fit timerange,
  (7) references for the X-ray and optical decay slopes. $^a$ The
  X-ray decay index for GRB\,970508 is approximate since there was
  substantial temporal and spectral variability for this burst. $^b$
  Based on the findings of Berger, Frail \& Kulkarni
  (2001)\nocite{bkf01}, we subtracted a constant component for the
  host galaxy of this burst. $^c$ \citet{bdf+01} find evidence for a
  jet break at t$\sim$26 d based on radio and late-time optical
  measurements. A fit to the data with a standard jet model in a
  constant density medium gives reasonable solutions for
  $\alpha_o\simeq 2.4$.} 
\tablerefs{(1) \citet{cfh+97}; (2) \citet{rei99}; (3) \citet{fpg+99};
  (4) \citet{pcf+99}; (5) \citet{sps+01}; (6) \citet{ddc+98}; (7)
  \citet{iaa+98}; (8) \citet{rlm+99}; (9) \citet{naa+99}; (10)
  \citet{hkp+99}; (11) \citet{vgo+99}; (12) \citet{kdo+99}; (13)
  \citet{psa+01}; (14) \citet{hbf+99}; (15) \citet{smp+00}; (16)
  \citet{fbg+00}; (17) \citet{hum+00}; (18) \citet{jfg+01}; (19)
  \citet{bdf+01}; (20) \citet{pbk+02}; (21) \citet{pgg+01}; (22)
  \citet{hys+01}; (23) \citet{zka+01}; (24) \citet{grb+03}.}
\end{deluxetable}

\clearpage

\clearpage

%\clearpage
%\begin{figure}
%\epsscale{0.8}
%\plotone{six_panel.ps}
%\caption{Radio light curves of six GRBs at a frequency of 8.46
%  GHz. 
%\label{fig:six}}
%\end{figure}

\end{document}